# Hidden Symmetry of Flexoelectric Coupling

Eugene A. Eliseev[1] and Anna N. Morozovska[2,3*]

[1] *Institute for Problems of Materials Science, National Academy of Sciences of Ukraine, 3, Krjijanovskogo, 03142 Kyiv, Ukraine*

[2] *Institute of Physics, National Academy of Sciences of Ukraine, 46, Prospekt Nauky, 03028 Kyiv, Ukraine*

[3] *Bogolyubov Institute for Theoretical Physics, National Academy of Sciences of Ukraine, 14-b Metrolohichna str. 03680 Kyiv, Ukraine*

**Abstract**

Considering the importance of the flexoelectric coupling for the physical understanding of the gradient-driven couplings in mesoscale and nanoscale solids, one has to determine its full symmetry and numerical values. The totality of available experimental and theoretical information about the flexocoupling tensor symmetry (specifically the amount of measurable independent components) and numerical values is contradictory. However, the discrepancy between the theory and experiment can be eliminated by consideration all possible inner symmetries of the flexocoupling tensor and physical limits on it components values. Specifically, this study reveals the inner "hidden" symmetry of the static flexoelectric tensor that allows minimizing the number of its independent components. Revealed hidden symmetry leads to nontrivial physical sequences, namely it affects on the upper limits of the static flexocoupling constants. Also we analyze the dynamic flexoelectric coupling symmetry and established the upper limits for the numerical values of its components. These results can help to understand and quantify the fundamentals of the gradient-type couplings in different solids.

[*] anna.n.morozovska@gmail.com

# I. INTRODUCTION

The flexoelectric coupling is principally important for fundamental insight into the complex electromechanics of meso-, nanoscale inorganic solids and biological systems, for which the strong strain gradients are inevitable present at the surfaces, interfaces, around point and topological defects [1, 2, 3, 4, 5, 6, 7].

**Static flexoelectric coupling.** The static flexoelectric effect is the appearance of elastic stress $\sigma_{ij}$ in response to electric polarization gradient $\partial P_k/\partial x_l$ (direct effect), and, vice versa, the polarization $P_i$ appears as a response to the strain gradient $\partial u_{ij}/\partial x_l$ (inverse effect) [8, 9, 10]. Variation of corresponding Lifshitz invariant included to the free energy functional of a solid [11],

$$F_{LI} = \int_V d^3 r \frac{f_{ijkl}}{2}\left(P_k \frac{\partial u_{ij}}{\partial x_l} - u_{ij} \frac{\partial P_k}{\partial x_l}\right), \tag{1a}$$

leads to the linear relations,

$$\sigma_{ij} = f_{ijkl}\frac{\partial P_k}{\partial x_l}, \qquad P_k = f_{ijkl}\frac{\partial u_{ij}}{\partial x_l}. \tag{1b}$$

All components of the flexocoupling tensor $f_{ijkl}$ are symmetrical with respect to the first pair of indices, $f_{ijkl} = f_{jikl}$, and the trivial fact originates from the internal symmetry of the strain tensor, $u_{ij} \equiv u_{ji}$, that in turn follows from its definition in continuum media approximation, $u_{ij} = (\partial U_i/\partial x_j + \partial U_j/\partial x_i)/2$, via displacement vector components $U_i$. No other relations between its components follow from Eqs.(1).

Notably, the static flexoelectric effect is allowed by symmetry in all 32 crystalline point groups, because the strain gradient breaks the inversion symmetry, making the static flexoeffect omnipresent. For instance, Shu et al. [12], Quang and He [13], consider possible symmetries of the flexoelectric tensor and derived the number of its independent components for each symmetry. Let us underline the fact that Shu et al., Quang and He used the invariance of the flexocoupling tensor $f_{ijkl}$ to the permutation of the first one pair of indices, $f_{ijkl} = f_{jikl}$. This work is about another nontrivial fact, that $f_{ijkl} = f_{ilkj}$ (the derivation is given below in **Section II**). Hence both the trivial ( $f_{ijkl} \equiv f_{jikl}$ ) and nontrivial ( $f_{ijkl} = f_{ilkj}$ ) equalities are valid.

It has been shown that the static flexocoupling induces imprint [14, 15, 16], internal bias [17, 18] and dead layer effect [19, 20] in ferroelectric thin films, and vortices in superlattices [21]. Flexoelectric coupling strongly changes the structure and electro-transport properties of the domain walls and interfaces in ferroelectrics [22, 23, 24, 25] and ferroelastics [26, 27, 28]. The flexocoupling leads to the hardening of solids at nano-indentation [29, 30, 31], significantly affects on the local electrochemical strains appeared in materials with mobile charges [32, 33], as well as on the mechanical writing of ferroelectric polarization by the tip [34]. The flexocoupling can induce incommensurate spatially modulated phases in many ferroics including antiferroelectric and antiferrodistortive ones [28, 35, 36,37]. Since a wave excitation of any nature is impossible without a local gradient of the corresponding physical quantity [38, 39], the flexoelectric effects significantly influence the propagation of surface acoustic waves in non-piezoelectric solids [40].

Considering the importance of the flexoelectricity for the physical understanding of the gradient-driven mesoscale and nanoscale couplings in solids, one has to determine the *internal symmetry* and *numerical values* of static flexocoupling tensor $f_{ijkl}$.

**Numerical values of the static flexocoupling tensor.** Typically the values $f_{ijkl}$ calculated from the first principles [41, 42, 43, 44, 45] can be several orders of magnitude smaller than those measured experimentally [46, 47, 48, 49]. The discrepancy motivated Yudin, Ahluwalia and Tagantsev [50] to establish theoretically the upper limits for the values of the static flexoelectric coefficients $f_{ijkl}$ in solids with cubic parent phase symmetry. The calculated maximal values of $f_{ijkl}$ showed that the anomalously high flexoelectric coefficients measured for perovskite ceramics [46, 47, 48] cannot be related with the manifestation of the static flexoelectric effect. Morozovska et al. [51, 52, 53] established that spatially modulated phases appear and become stable in commensurate ferroelectrics if the flexocoupling constants exceed the maximal critical values, which depend on the electrostriction and elastic constants, temperature, and gradient coefficients in the Landau-Ginzburg-Devonshire functional. Hence the comparison of the aforementioned experimental and theoretical results tell us that numerical values of the static flexocoupling tensor are rather contradictory, indicating on a limited understanding of the effect strength.

## II. HIDDEN SYMMETRY OF THE FLEXOCOUPLING TENSOR

Specifically, for one of the highest m3m cubic point symmetry group three independent components of the flexocoupling tensor are $f_{1111} = f_{2222} = f_{3333}$, $f_{1122} = f_{1133} = f_{2233} = f_{3322} = ...$ and the third component $f_{1212} = f_{1313} = f_{2323} = f_{2121} = ...$ that is identically equal to $f_{1221} = f_{1331} = f_{2332} = f_{2112} = ...$ due to the permutation symmetry, $f_{ijkl} \equiv f_{jikl}$.

At first Zubko et al. [49] measured experimentally the tensor of flexoelectric effect in SrTiO$_3$ with cubic parent phase m3m symmetry and stated that it includes all three components, $f_{1111}$, $f_{1122}$ and $f_{1212}$. The flexoelectric coefficients measurement [49] is the dynamical bending of thin electrode plates using "three knives" setup with simultaneous measurement of a displacement current. The applied stain has zero average value and changes its sign inside the plate (i.e. distribution with "pure gradient"). The induced polarization is proportional to the strain gradient value determined from the geometry of deformed system.

However, later Zubko et al. [54] recognized that it was mathematically impossible to define all three components of the flexocoupling tensor from the quasi-static bending of plates [49] alone and suggested to use the results of independent dynamical measurements based on Brillouin scattering by Hehlen et al.[55] (see also the work of Tagantsev et al. [56]). Specifically, Zubko et al. wrote in the erratum [54] to their earlier paper [49] that "despite the use of different sample geometries and crystallographic orientations, only *two independent equations* involving $f_{1111}$, $f_{1122}$, and $f_{1212}$ can be obtained from bending experiments alone, and additional information is required to find the individual tensor components." At the same time additional equation for $f_{1212}$ was obtained from independently measured value based on Brillouin scattering data [55]. It should be noted that obtained in this way values of $f_{1122}$ and $f_{1212}$ appeared close to each other (7 nC/m and 5.8 nC/m respectively), especially in comparison with $f_{1111}$=0.2 nC/m.

It should be also noted that Yudin et al. [50] found only two independent conditions for upper limits for three independent flexoelectric tensor components $f_{1111}$, $f_{1122}$, and $f_{1212}$ required for the homogeneous phase stability under the absence of higher elastic gradients. From the phenomenological theory of phonon dispersion relations (see e.g. [51]) not all the components of static flexocoupling tensor are involved. Hong and Vanderbilt [43] calculated flexoelectric coefficients under the *assumption* that "transversal" flexocoupling coefficient, $f_T = f_{1122} - f_{1212}$ is zero. Below we will prove the assumption $f_{1122} - f_{1212} \equiv 0$ by exploring the "hidden" symmetry of the flexocoupling.

Hence, the primary goal of this work is to establish the inner "hidden" symmetry of the static flexocoupling tensor that allows minimizing the number of its independent components. Actually, the integration in parts for in Eq.(1a) for a *bulk infinite solid* yields:

$$\int_{V\to\infty} d^3r \frac{f_{ijkl}}{2}\left(u_{ij}\frac{\partial P_k}{\partial x_l} - P_k\frac{\partial u_{ij}}{\partial x_l}\right) = \int_{V\to\infty} d^3r f_{ijkl} u_{ij}\frac{\partial P_k}{\partial x_l} = ... = \\ \int_{V\to\infty} d^3r \left(\frac{f_{ijkl} + f_{jikl}}{2}\right)\frac{\partial U_i}{\partial x_j}\frac{\partial P_k}{\partial x_l} = ... = \int_{V\to\infty} d^3r \left(\frac{f_{ilkj} + f_{likj}}{2}\right)\frac{\partial U_i}{\partial x_j}\frac{\partial P_k}{\partial x_l}$$ (2)

Detailed derivation of each step in Eq.(2) is given in **Appendix A,** where we dropped all surface integrals (e.g. $-\int_S d^2r \frac{f_{ijkl}}{2} u_{ij} P_k n_l$), regarding them negligible for an infinite bulk material in a continuum media approximation. This is correct, because the flexoelectric tensor $f_{ijkl}$ describes the macroscopic bulk properties of the material in a continuum media approximation and therefore its symmetry cannot depend on the surface surrounding the sample.

Since the summation in Eq.(2) is performed on the "dumb" indices, they could be redefined in arbitrary manner (see **Appendix A**), so that one concludes that the latter equality in Eq.(2) is possible for the arbitrary polarization and displacement only if

$$f_{ijkl} = f_{ilkj}. \qquad (3)$$

Nontrivial relation Eq.(3) along with the trivial relation $f_{ijkl} = f_{jikl}$ lead to the conclusion that only *two* components $f_{1111}$ and $f_{1122} = f_{1221} = f_{1212}$ are *independent* for m3m symmetry (instead of three or four components as regarded previously).

Note, that for the isotropic media only one non-trivial component of tensor $f_{ijkl}$ is possible in contrast to findings of Le Quang and He [13], who did consider additional internal symmetry of flexoelectric tensor.

That say Eq.(3) is a manifestation of the static flexoeffect "hidden" symmetry. The conclusion can explain previously unexplained conclusion from Zubko et al experiments [49] and Hong and Vanderbilt *ab initio* calculations [43] for materials with cubic m3m symmetry. Specifically for m3m the effective flexoelectric coefficients $f'_{1111}(\varphi) = f_{1111} - \mu \sin^2(2\varphi)$ and $f'_{1212}(\varphi) = f_{1212} + \mu \sin^2(2\varphi)$ can be measured experimentally as a function of sample rotation angle $\varphi$. The anisotropy value $\mu = (f_{1111} - f_{1122} - 2f_{1212})/2$ is equal to $(f_{1111} - 3f_{1212})/2$ because $f_{1122} = f_{1212}$ accordingly to Eq.(3). From these relations the "true" flexocoefficients are only $f_{1111} = f'_{1111}(0)$ and $f_{1212} = f'_{1212}(0)$. Since $f'_{1212}(\pi/4) \equiv (f_{1111} - f_{1212})/2$ we obtain that $2f'_{1212}(\pi/4) \equiv f'_{1111}(0) - f'_{1212}(0)$ as the direct consequence of Eq.(3), and the latter relation can be verified experimentally. The angular dependences $f'_{1111}(\varphi)$ and $f'_{1212}(\varphi)$ are shown in **Fig. 1.**

Since relation (3) is based on the transformation properties of the phenomenological free energy of the bulk material in a continuum media approximation, it has nothing similar with e.g. (semi)-microscopic theories [57], predicting Cauchy relations, $c_{ijkl} = c_{jikl}$, for elastic stiffness $c_{ijkl}$ [58].

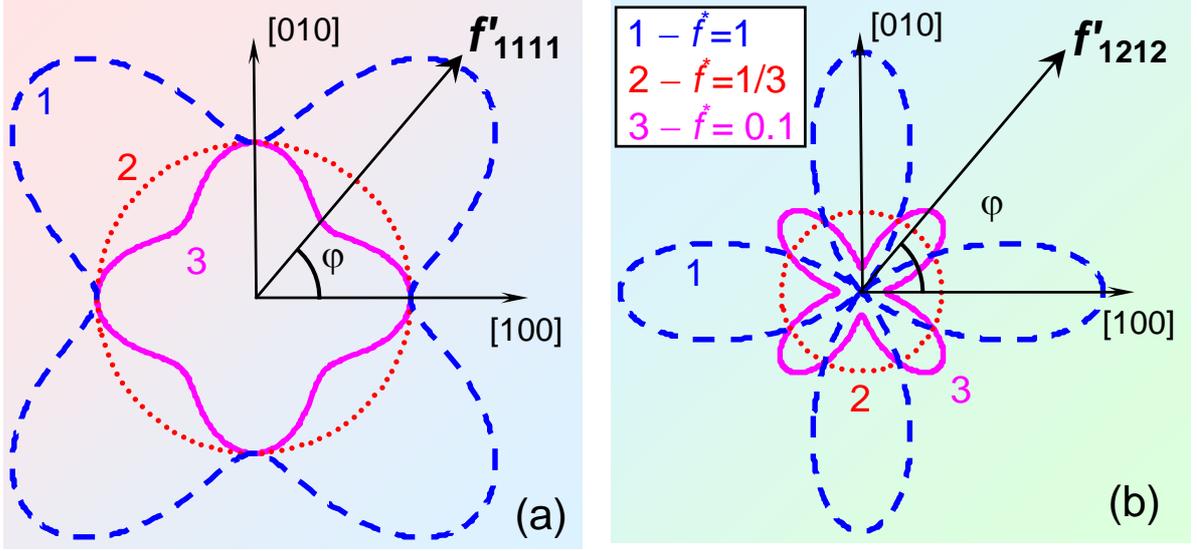

**FIG. 1.** Effective flexoelectric coefficient $f'_{1111}(\varphi) = f_{1111} - \mu\sin^2(2\varphi)$ **(a)** and $f'_{1212}(\varphi) = f_{1212} + \mu\sin^2(2\varphi)$ **(b)** as a function of the sample rotation angle $\varphi$ calculated for several values of the ratio $f^* = f_{1212}/f_{1111}$ equal to 1 (dashed blue curves 1), 1/3 (dotted red curves 2) and 0.1 (solid magenta curves 3).

For symmetries lower than cubic the revealed "hidden" symmetry described by Eq.(3) leads to nontrivial physical sequences, namely it should affect on the upper limits of the static flexocoupling constants established by Yudin et al [50]. To illustrate the statement let us analyze the expression derived in Ref.[33] for the Fourier spectra of linear static dielectric susceptibility $\chi_{ij}(\mathbf{r})$. The expression for inverse susceptibility, $\tilde{\chi}_{ij}^{-1}(\mathbf{k})$, valid for a dielectric solid (or in a paraelectric phase of ferroelectric), has a relatively simple form:

$$\tilde{\chi}_{ij}^{-1}(\mathbf{k}) = 2\alpha\delta_{ij} + \left(g_{ipjl} - f_{miln}f_{i'pjj'}k_n k_{j'}S_{mi'}(\mathbf{k})\right)k_p k_l . \qquad (4)$$

The summation in Eq.(4) is performed over all repeating indexes allowing for Eq.(3). The coefficient $\alpha$ is positive and temperature dependent in the paraelectric phase of ferroelectrics as $\alpha = \alpha_T(T - T_C)$. $g_{ijkl}$ is a positively defined symmetrized tensor of polarization gradient coefficients [33]. Inverse matrix is $S_{ik}^{-1}(\mathbf{k}) = c_{ijkl}k_l k_j$ and elastic compliances permutative symmetry is $c_{ijkl} = c_{jikl} = c_{ijlk}$. The equations $\left[g_{ipjl} - f_{mlin}f_{i'pjj'}k_n k_{j'}S_{mi'}(\mathbf{k})\right]k_p k_l = 0$ ($i, j = 1, 2, 3$) define zeroes $\tilde{\chi}_{ij}^{-1}(\mathbf{k}) = 0$ at $\alpha = 0$. For the wave vector $\mathbf{k} = (0, 0, k)$ directed along direction $[0, 0, 1]$ the matrix $S_{ik}^{-1}(\mathbf{k}) = c_{i3k3}k^2$ and so $S_{mi'}(\mathbf{k}) = \tilde{s}_{mi'}^{(3)}/k^2$ at that $c_{i3n3}\tilde{s}_{nm}^{(k)} = \delta_{im}$. In this way the stability conditions for $[1, 0, 0]$, $[0, 1, 0]$ and $[0, 0, 1]$ directions of the wave vector acquire the form

$$f_{mikk}f_{ljkk}\tilde{s}_{ml}^{(k)} \leq g_{ikjk} \qquad \text{(without summation over } k = 1, 2, 3 \text{)} \qquad (5)$$

and define the upper limits of the static flexocoupling constants required for the homogeneous state stability for arbitrary point symmetry.

The explicit form of the stability conditions for the flexoelectric coefficients depends on the direction of the wave vector of the fluctuations and concrete symmetry. In order to show the role of the flexocoupling hidden symmetry we derived the stability conditions imposed on $f_{ijkl}$ values for m3m symmetry, transverse (TO) and longitudinal (LO) optic modes propagation in [100], [110] and [111] directions. Results are listed them in the second column of **Table I**. For m3m symmetry Yudin et al [50] considered [100], [110] and [111] **k**-directions for TO mode (see the the third column of **Table I**).

For the direction [100] and [110], our conditions for TO mode coincide with the conditions obtained by Yudin et al. For the direction [111] our condition has more simple form than the one derived by Yudin et al, since the additional constraint, $f_{1212} \equiv f_{1122}$, is valid accordingly to Eq.(3).

**Table 1.** Comparative table

| Wave vector direction | This work gives the following relations for m3m cubic symmetry in matrix notations | Yudin et al. [50] obtained relations for m3m cubic symmetry in matrix notations |
|---|---|---|
| **[100]** | **TO-mode:** $f_{1212}^2 \leq c_{1212} g_{1212}$ | **TO-mode:** $f_{1212}^2 \leq c_{1212} g_{1212}$ |
| | **LO-mode:** $f_{1111}^2 \leq c_{1111} g_{1111}$ | **LO-mode** was not considered |
| **[110]** | **TO-mode:** $(f_{1111} - f_{1122})^2 \leq (c_{1111} - c_{1122})(g_{1111} - g_{1122})$ | **TO-mode** $(f_{1111} - f_{1122})^2 \leq (c_{1111} - c_{1122})(g_{1111} - g_{1122})$ |
| | **LO-mode** $(f_{1111} + 3f_{1212})^2 \leq (c_{1111} + c_{1122} + 2c_{1212})(g_{1111} + g_{1122} + 2g_{1212})$ | **LO-mode** was not considered |
| **[111]** | **TO-mode** $f_{1111}^2 \leq (c_{1111} - c_{1122} + c_{1212})(g_{1111} - g_{1122} + g_{1212})$ | **TO-mode** $(f_{1111} - f_{1122} + f_{1212})^2 \leq (c_{1111} - c_{1122} + c_{1212})(g_{1111} - g_{1122} + g_{1212})$ |
| | **LO-mode** $(f_{1111} + 6f_{1122})^2 \leq (c_{1111} + 2c_{1122} + 4c_{1212})(g_{1111} + 2g_{1122} + 4g_{1212})$ | **LO-mode** was not considered |

### III. INNER SYMMETRY AND UPPER LIMITS OF DYNAMIC FLEXOELECTRIC EFFECT

From considerations of the symmetry theory stating that all terms and invariants, which existence does not violate the symmetry of the system, are allowed, Tagantsev et al. (see reviews [2, 3] and refs therein) predicted the existence of *dynamic flexoelectric effect* originated from the cross-term in the kinetic energy, proportional to polarization components $P_i$ and elastic displacement $U_j$ time derivatives,

$$M_{ij}\frac{\partial P_i}{\partial t}\frac{\partial U_j}{\partial t}, \qquad (6)$$

where $M_{ij}$ is a flexodynamic tensor, much earlier considered microscopically by Shirane et al [59]. The dynamic flexoelectric effect corresponds to the polarization response to accelerated motion of the medium in the time domain, $P_i = -\frac{M_{ij}}{\alpha}\frac{\partial^2 U_j}{\partial t^2}$, where α is the dielectric stiffness.

At present the situation with the numerical values $M_{ij}$ of dynamic flexoeffect is more complex and controversial that for the static one, because there are microscopic theories in which the effect is absent (see e.g. [60]). Namely, Stengel [44] has shown that since both polarization and elastic displacement are supposed to be normal modes of the crystal at the center of Brillouin zone, they should diagonalize the dynamical matrix leading to the absence of the cross-term in the matrix. It should be noted that the Stengel result [44] was argued later on by Kvasov and Tagantsev [61], who evaluated the strength of the dynamic flexoelectric effect in SrTiO$_3$ from microscopic calculations and it appeared comparable to that of the static bulk flexoelectric effect. Note that the best fitting of the soft phonon spectra observed in SrTiO$_3$, PbTiO$_3$, Sn$_2$P$_2$(S,Se)$_6$ performed by Morozovska et al [51, 52] corresponds to nonzero values of dynamic flexoelectric coefficient, which are of the same order as the one calculated by Kvasov and Tagantsev [61].

Since the indirect evidences followed from the fitting with many parameters [51, 52], the next goal of this work is to proof that the critical values of dynamic flexocoupling tensor should exist. For this let us analyze the expression derived in Ref.[33] for the linear dynamic dielectric susceptibility $\chi_{ij}(t,\mathbf{r})$ of a dielectric solid (or of a ferroelectric in a paraelectric phase). The expression for inverse susceptibility has a relatively simple form its Fourier image in frequency-wave vector space $(\omega,\mathbf{k})$, and allowing for Eq.(3) it reads:

$$\tilde{\chi}_{ij}^{-1}(\omega,\mathbf{k}) = (2\alpha - \mu\omega^2)\delta_{ij} + g_{ipjl}k_p k_l - (f_{miln}k_n k_l - M_{mi}\omega^2)(f_{i'jpj'}k_j k_p - M_{i'j}\omega^2)S_{mi'}(\mathbf{k},\omega) \qquad (7)$$

The summation in Eq.(7) is performed over all repeating indexes. Inverse matrix $S_{ik}^{-1}(\mathbf{k},\omega) = c_{ijkl}k_l k_j - \rho\omega^2\delta_{ik}$ ($\delta_{ij}$ is the Kroneker symbol), ρ is the density of a solid and μ is a kinetic coefficient. The coefficient α is positive for dielectrics and is temperature dependent in the paraelectric phase of ferroelectrics (e.g. $\alpha = \alpha_T(T - T_C)$ in the vicinity of Curie temperature $T_C$), $g_{ijkl}$ is a positively defined symmetrized tensor of polarization gradient coefficients [33]. The solutions of characteristic equation $\tilde{\chi}_{ij}^{-1}(\mathbf{k},\omega) = 0$ give the optic and acoustic soft phonon dispersion law in the parent phase of ferroics. Putting condition $\mathbf{k} = 0$ in Eq.(7) we obtain the equations $2\alpha\delta_{ij} + \left(\dfrac{M_{mi}M_{mj}}{\rho} - \mu\delta_{ij}\right)\omega^2 = 0$. The nontrivial solutions of these equations for $\alpha \geq 0$ correspond to the values of optic mode at $\mathbf{k} = 0$,

$$\omega_i^{opt}(\mathbf{k}=0) = \sqrt{\dfrac{2\alpha(T)}{\mu - \left(M_{1i}^2 + M_{2i}^2 + M_{3i}^2\right)/\rho}}, \qquad (i=1, 2, 3). \qquad (8)$$

For diagonal tensor $M_{ij} = M_{ii}\delta_{ij}$ the solution (8) exists under the conditions

$$M_{ii}^2 < \mu\rho \quad \text{(without summation over } i=1, 2, 3). \qquad (9)$$

For cubic m3m symmetry the dynamic flexocoupling tensor is diagonal, $M_{11} = M_{22} = M_{33} = M$, and the inequality $M^2 < \mu\rho$ follows from Eq.(9). Numerical estimates give $|M_{11}| < 11.3\times10^{-8}$ V s$^2$/m$^2$ for $\mu$ =1.59×10$^{-18}$ s$^2$Jm, $\rho$=7.986×10$^3$ kg/m$^3$ (corresponding to PbTiO$_3$ in a cubic paraelectric phase at $T$=783 K, $T_C$=752 K); and $|M_{11}| < 32.9\times10^{-8}$ V s$^2$/m$^2$ for $\mu$=22×10$^{-18}$ s$^2$Jm, $\rho$=4.930×10$^3$ kg/m$^3$ (corresponding to paraelectric SrTiO$_3$ at $T$=120 K). The values of $|M_{11}|$ corresponding to the best fitting of experimentally measured soft phonon spectra appeared significantly lower, namely $|M_{11}| = 2\times10^{-8}$ V s$^2$/m$^2$ for PbTiO$_3$ and $|M_{11}| = 22\times10^{-8}$ V s$^2$/m$^2$ for SrTiO$_3$ [33].

The temperature dependence of expression (8) can serve for determination of dynamic flexocoupling value [compare curves 1-4 in **Fig. 2(a)**], because any deviation of the ratio $\eta = \mu(\omega^{opt})^2/2\alpha_T(T-T_C)$ from unity at $T > T_C$ can follow from the existence of the temperature-dependent term in denominator, $\left(M_{1i}^2 + M_{2i}^2 + M_{3i}^2\right)/\rho$. Almost horizontal solid lines 2, 3 and 4 in **Fig. 2(b)** correspond to $\eta$ values calculated for different $M_{ij} \neq 0$. They essentially differ from $\eta = 1$ (dashed line 1) calculated for $M_{ij} = 0$. Also we regard that $\mu$ and $M_{ii}$ are temperature independent, $\alpha = \alpha_T(T-T_C)$ and the density $\rho = \rho_T[1+\beta_T(T_C-T)]$ obeys linear thermal expansion law; $\beta_T$ is the thermal expansion coefficient.

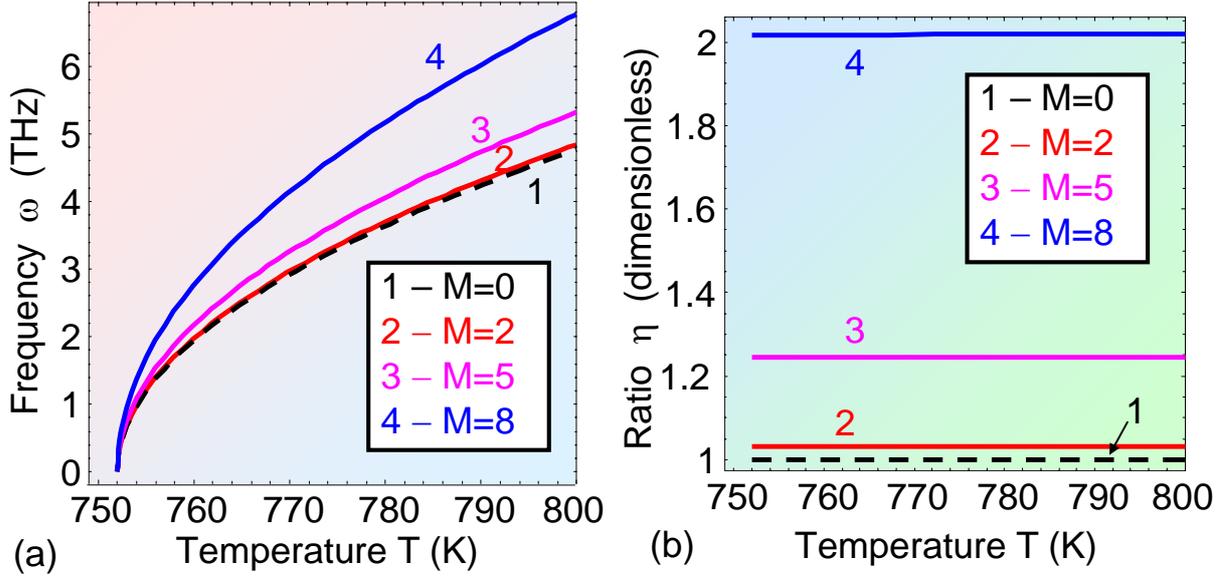

**FIG. 2.** Temperature dependence of the soft phonon TO frequency $\omega^{opt}$ **(a)** and the ratio **(b)** $\eta = \mu(\omega^{opt})^2 / 2\alpha_T(T-T_C)$ calculated from Eq.(8) at $M_{ij}=0$ (dashed black curve) and $|M_{ii}|=(2, 5, 8)\times 10^{-8}$ V s$^2$/m$^2$ (solid red, magenta and blue curves), $\alpha_T = 3.765\times 10^5$ F/m, $T_C$=752 K, $\mu$ =1.59×10$^{-18}$ s$^2$Jm, $\rho$=7.986×10$^3$ kg/m$^3$ and $\beta_T = 3.2\times 10^{-5}$ K$^{-1}$ corresponding to PbTiO$_3$ in a paraelectric phase.

## IV. CONCLUSION

To resume, this study revealed the inner "hidden" symmetry of the static flexoelectric tensor that allows minimizing the number of its independent components. Revealed hidden symmetry leads to nontrivial physical sequences, namely it affects on the upper limits of the static flexocoupling constants. Also we analyze the dynamic flexoelectric coupling symmetry and established the upper limits for the numerical values of its components. These results can help to understand and quantify the fundamentals of the gradient-type flexocoupling.

**ACKNOWLEDGMENTS.** E.A.E. and A.N.M. contributed equally to the paper.

## APPENDIX A. Derivation of relation (3)

Let us consider the contribution of flexoelectric coupling to the free energy

$$\int_{V\to\infty} d^3r \frac{f_{ijkl}}{2}\left(u_{ij}\frac{\partial P_k}{\partial x_l} - P_k \frac{\partial u_{ij}}{\partial x_l}\right) = \begin{vmatrix}\text{integration in parts}\\ \text{in the second integral}\end{vmatrix} = \int_{V\to\infty} d^3r f_{ijkl} u_{ij}\frac{\partial P_k}{\partial x_l} \quad (A.1)$$

When integrating in parts in Eq.(A.1) we neglected the surface integrals like $-\int_S d^2r \frac{f_{ijkl}}{2} u_{ij} P_k n_l$ hereinafter, since it can be done for a bulk material in a continuum media approximation. Since $u_{ij} = (\partial U_i/\partial x_j + \partial U_j/\partial x_i)/2$, the transformations of Eq.(A.1) yields

$$\int_{V\to\infty} d^3r \frac{f_{ijkl}}{2}\left(\frac{\partial U_i}{\partial x_j}+\frac{\partial U_j}{\partial x_i}\right)\frac{\partial P_k}{\partial x_l} = \begin{vmatrix}\text{renaming indices}\\ "i"\to"\tilde{j}" \text{ and } "j"\to"\tilde{i}"\\ \text{in the second term}\end{vmatrix} = \int_{V\to\infty} d^3r \left(\frac{f_{ijkl}}{2}\frac{\partial U_i}{\partial x_j}+\frac{f_{\tilde{j}\tilde{i}kl}}{2}\frac{\partial U_{\tilde{i}}}{\partial x_{\tilde{j}}}\right)\frac{\partial P_k}{\partial x_l} =$$

$$=\begin{vmatrix}\text{renaming indices}\\ "\tilde{j}"\to"j" \text{ and } "\tilde{i}"\to"i"\\ \text{in the second term}\end{vmatrix} = \int_{V\to\infty} d^3r \left(\frac{f_{ijkl}+f_{jikl}}{2}\right)\frac{\partial U_i}{\partial x_j}\frac{\partial P_k}{\partial x_l} =$$

$$=\begin{vmatrix}\text{integration in parts}\\ \text{for both multipliers } \frac{\partial U_i}{\partial x_j} \text{ and } \frac{\partial P_k}{\partial x_l}\end{vmatrix} = \int_{V\to\infty} d^3r \left(\frac{f_{ijkl}+f_{jikl}}{2}\right)\frac{\partial U_i}{\partial x_l}\frac{\partial P_k}{\partial x_j} = \quad\text{(A.2)}$$

$$=\begin{vmatrix}\text{renaming indices}\\ "l"\to"\tilde{j}" \text{ and } "j"\to"\tilde{l}"\\ \text{in both terms}\end{vmatrix} = \int_{V\to\infty} d^3r \left(\frac{f_{i\tilde{l}k\tilde{j}}+f_{\tilde{l}ik\tilde{j}}}{2}\right)\frac{\partial U_i}{\partial x_{\tilde{j}}}\frac{\partial P_k}{\partial x_{\tilde{l}}} = \begin{vmatrix}\text{omit}\\ \text{tilda}\end{vmatrix} = \int_{V\to\infty} d^3r \left(\frac{f_{ilkj}+f_{likj}}{2}\right)\frac{\partial U_i}{\partial x_j}\frac{\partial P_k}{\partial x_l}$$

Comparing underlined by "blue" and "red" steps of the transformations (A.2) for an arbitrary coordinate-dependent function $\frac{\partial U_i}{\partial x_j}\frac{\partial P_k}{\partial x_l}$, one leads to the relation

$$f_{ijkl} + f_{jikl} = f_{ilkj} + f_{likj}, \quad\text{(A.3)}$$

since $f_{ijkl} \equiv f_{jikl}$ and $f_{ilkj} \equiv f_{likj}$, due to the symmetry of the strain tensor, $u_{ij} \equiv u_{ji}$, Eq.(A.3) elementary leads to Eq.(3):

$$f_{ijkl} = f_{jikl} = f_{ilkj} = f_{likj} \quad\text{(A.4)}$$

Hence we proved that besides the relation $f_{ijkl} \equiv f_{jikl}$ there is one more relation $f_{ijkl} = f_{ilkj}$, Eq. (3), that imposes additional constrains on the structure of the flexoelectric tensor.